\documentclass[letter,twocolumn]{jpsj3} %% for letters
\def\runauthor{Takuya \textsc{Iizuka} \textit{et al}.}
\title{
Temperature- and Magnetic-Field-Dependent Optical Properties of Heavy Quasiparticles in YbIr$_2$Si$_2$.
}
\author{
Takuya \textsc{Iizuka}$^1$,
Shin-ichi \textsc{Kimura}$^{2,1,}$\thanks{E-mail address: kimura@ims.ac.jp}, 
Alexander \textsc{Herzog}$^3$,
J\"org \textsc{Sichelschmidt}$^3$,
Cornelius \textsc{Krellner}$^3$,
Christoph \textsc{Geibel}$^3$,
and
Frank \textsc{Steglich}$^3$
}
\inst{
$^1$School of Physical Sciences, The Graduate University for Advanced Studies (SOKENDAI), Okazaki 444-8585 \\
$^2$UVSOR Facility, Institute for Molecular Science, Okazaki 444-8585 \\
$^3$Max Planck Institute for Chemical Physics of Solids, N\"othnitzer Stra\ss e 40, 01187 Dresden, Germany
}
\recdate{
\today
}
\abst{
We report the temperature- and magnetic-field-dependent optical conductivity spectra of the heavy electron metal YbIr$_2$Si$_2$. 
Upon cooling below the Kondo temperature ($T_{\rm K}$), we observed a typical charge dynamics that is expected for a formation of a coherent heavy quasiparticle state. 
We obtained a good fitting of the Drude weight of the heavy quasiparticles by applying a modified Drude formula with a photon energy dependence of the quasiparticle scattering rate that shows a similar power-law behavior as the temperature dependence of the electrical resistivity. 
By applying a magnetic field of 6~T below $T_{\rm K}$, we found a weakening of the effective dynamical mass enhancement by about 12~\% in agreement with the expected decrease of the $4f$-conduction electron hybridization on magnetic field.
}
\kword{heavy fermion, YbIr$_2$Si$_2$, optical conductivity, electronic structure}
\begin{document}
\maketitle
%
%%%%%%%%%%%%%%%%%%%%%%%%%%%%%%%%%%%%%%%%%%%%%%%%%%%%%%%%%%%%
%
%\section{Introduction}
%
The charge dynamics of heavy electrons in cerium (Ce) and ytterbium (Yb) intermetallic compounds are strongly determined by the physics of a Kondo lattice, which, in particular, may be manifested in quantum critical phenomena such as a pronounced deviation from the renormalized Landau-Fermi liquid (LFL) behavior.\cite{Gegenwart2008} 
In this respect, the interplay between itinerant heavy electron states with non-magnetic ground states and local magnetic ordering ground states at low temperature is attracting attention because the heavy electrons originate from the $c$-$f$ hybridization between conduction electrons and local $4f$ states located near the Fermi level ($E_{\rm F}$).~\cite{Hewson1993}

In the highly itinerant region, the system of heavy electrons is regarded as a renormalized LFL, {\it i.e.,} the electron-electron scattering rate as well as the electrical resistivity is proportional to $T^2$. 
However, near the boundary between the itinerant and the magnetically ordered states, the physical properties show a so-called non-Fermi liquid (NFL) behavior being inconsistent with the expectation for a LFL state. 
In order to understand the origin of these new physical properties, investigations of the far-infrared optical properties provide essential information on the $c$-$f$ hybridization effect in the electronic structure.
Moreover, application of a magnetic field can strongly affect the low-energy excitations of a heavy electron state providing the opportunity to control the effect of the Kondo interaction on the effective electronic mass.~\cite{Dordevic2006}

The electrodynamical response of heavy electrons is characterized by a scattering rate
\begin{equation}
\frac{1}{\tau(\omega,T)}=\tau_0^{-1}+aT^n+b(\hbar\omega)^n,
\label{tau1}
\end{equation}
where $\tau_0$ is the residual life time, $a$ and $b$ denote proportional constants of temperature and photon energy, respectively.~\cite{DG}
A typical LFL behavior with $n=2$ common to the temperature- and energy-term is seen, for instance, in the heavy fermion materials CeAl$_3$ and YbAl$_3$.~\cite{Awasthi1993,Okamura2004} 
However, although a $n\neq2$ behavior has been observed in electrical resistivity measurements of many NFL compounds, there are only a few reports of a corresponding energy dependence of the scattering rate.~\cite{Stewart2001}
Among those, YbRh$_{2}$Si$_{2}$ is a prototypical example, for which a pronounced NFL behavior of the scattering rate with $n=1$ is seen as a function of both temperature and photon energies.~\cite{Kimura2006}

In this Letter, we discuss the NFL issue of the optical scattering rate of heavy quasiparticles in the homologous material YbIr$_{2}$Si$_{2}$ (I-type) with a Kondo energy scale $T_{\rm K}\simeq40$~K.
This is the first Yb-based heavy fermion metal which has a LFL ground state close to a quantum phase transition. 
At low temperatures, LFL behavior is clearly evidenced in the electronic specific heat (constant Sommerfeld coefficient $\gamma=0.37$~J/mol~K$^2$ below 0.4~K) and in the electrical resistivity that is proportional to $T^2$ below 0.2~K.~\cite{Hossain2005}  
The power law of the latter behavior changes from 2 to 1.3 in the temperature range of 0.2~--~4~K.
A magnetic field of 6~T reduces the $\gamma$-value by a factor of $1.5$ and therefore reduces the mass enhancement correspondingly. 
Moreover, $\gamma$ becomes constant already below $\approx 2$~K. 
Thus, we investigated the dynamical effective mass and $\hbar\omega$-dependent scattering rate of YbIr$_2$Si$_2$ by measuring the temperature- and magnetic-field-dependent reflectivity [$R(\omega)$] spectra. 
Below $T_{\rm K}$, we found a clear Drude-type structure which is typical for the optical response of heavy quasiparticles and which contains a scattering rate that depends on photon energy and temperature by the same power law.
The magnetic field dependence turns out to be fairly weak but measurable and is consistent with the expectation of a mean field theory.

%------------------------------------------------------------------------------
%
%\section{Experimental}
%
Near-normal incident reflectivity [$R(\omega)$] spectra were acquired in a very wide photon-energy region of 2~meV -- 30~eV to ensure an accurate Kramers-Kronig analysis (KKA) for obtaining the complex optical conductivity
\begin{equation}
\hat{\sigma}(\omega)=\sigma_1(\omega)-i\sigma_2(\omega)=\frac{Ne^2\tau}{m_b}\frac{1}{1+i\omega\tau}. 
\label{NormalDrude}
\end{equation}
Here, $\sigma_1(\omega)$ and $\sigma_2(\omega)$ are the real and imaginary parts of the conductivity spectra, $N$ the effective charge-carrier density, $e$ the elementary charge, and $m_b$ the unrenormalized band mass at $T\gg T_{\rm K}$.
For the analysis, we also used spectra of the loss function $-\rm{Im}[\hat{\varepsilon}(\omega)^{-1}]$ ($\hat{\varepsilon}(\omega)$: complex dielectric function) that were derived from $R(\omega)$ spectra by a KKA.

We investigated single crystalline samples with as-grown and polished sample surfaces perpendicular to the $c$-axis.
The sample preparation as well as the magnetic and transport properties have been described elsewhere.~\cite{Hossain2005}
Rapid-scan Fourier spectrometers of Martin-Puplett and Michelson type were used at photon energies of 3--30~meV and 0.01--1.5~eV, respectively, with a specially designed feed-back positioning system to maintain the overall uncertainty level less than $\pm$0.2~\% at sample temperatures between 0.4--300~K using a $^4$He (down to 8~K) and a $^3$He (down to 0.4~K) cryostat.~\cite{Kimura2008}
In order to obtain accurate $R(\omega)$ spectra, reference spectra were measured by using the {\it in-situ} gold evaporated sample surface.
The magnetic field dependence was checked at $T=6$~K by applying one magnetic field of 6~T and in the photon energy range of 5--60~meV only.
At $T=300$~K, $R(\omega)$ was measured for energies 1.2--30~eV by using synchrotron radiation.~\cite{Fukui2001}
In order to obtain $\sigma_1(\omega)$ via a KKA of $R(\omega)$, the spectra were extrapolated below 2~meV with $R(\omega)=1-(2\omega/\pi \sigma_{DC})^{1/2}$, where $\sigma_{DC}$ is the direct current conductivity, and above 30~eV with a free-electron approximation, $R(\omega) \propto \omega^{-4}$.~\cite{DG}

To clarify the photon energy and temperature dependences of the quasiparticle scattering rate, an {\it extended} Drude analysis in terms of the effective mass [$m^*(\omega)$] and the scattering rate [$1/\tau(\omega)$] was performed.
The coherent part due to the underlying strong electron-electron correlations were treated by renormalized and frequency (photon energy) dependent $m^{*}(\omega)/m_{b}$ and $1/\tau(\omega)$;~\cite{DG}
\begin{equation}
\frac{m^{*}(\omega)}{m_b} = \frac{N e^2}{m_b \omega} \cdot Im\left(\frac{1}{\hat{\sigma}(\omega)}\right),
\end{equation}
\begin{equation}
\frac{1}{\tau(\omega)} = \frac{N e^2}{m_b} \cdot Re\left(\frac{1}{\hat{\sigma}(\omega)}\right).
\end{equation}
$N/m_b$ can be evaluated to be $7.1\times10^{21}$ cm$^{-3}$ by integrating the $\sigma_1(\omega)$ below the plasma edge of $\hbar\omega_p$~=~0.6~eV,
\begin{equation}
\frac{N}{m_b} = \frac{2}{\pi e^2}\int^{\omega_p}_{0} \sigma_1(\omega) d\omega .
\end{equation}

%------------------------------------------------------------------------------
%
%\section{Results and Discussion}
%
%%%%%%%%%%%%%%%%%  FIG.1  %%%%%%%%%%%%%%%%%%%%%%%%
\begin{figure}[t]
\begin{center}
\includegraphics[width=0.35\textwidth]{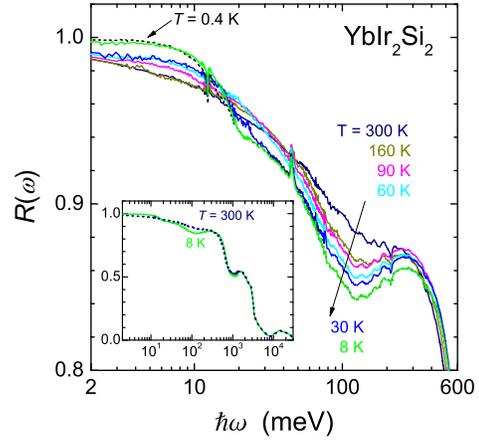}
\end{center}
\caption{
(Color online) Temperature dependence of the reflectivity spectrum $R(\omega)$ in the photon energy range of 2 -- 600~meV.
Inset: $R(\omega)$ at 8 and 300~K in the complete accessible range of photon energies up to 30~eV.
}
\label{Reflectivity}
\end{figure}
%%%%%%%%%%%%%%%%%%%%%%%%%%%%%%%%%%%%%%%%%%%%%
The measured $R(\omega)$ spectra are shown in Fig.~\ref{Reflectivity}. 
Upon cooling the sample the dip structure at $\hbar\omega$~=~110~meV becomes more pronounced and thereby bringing out a Drude-like reflectivity showing a remarkable increase at low energies. 
At 300~K, a dip structure appears at $\hbar\omega$~=~110~meV, but a Drude-like reflectivity, in which $R(\omega)$ increases with decreasing $\hbar\omega$, is realized.
With decreasing temperature, the dip structure at 110~meV becomes deeper due to the strong electron-electron correlations as observed in other Yb- and Ce-compounds.
For $T\le$30~K, a characteristic kink structure appears at 20~meV.
This happens below $T_{\rm K}$ and, therefore, the steep increase below this kink is probably part of the Drude response of the heavy quasiparticles.
Note that the peaks at 12 and 40~meV sharpen with decreasing temperature, indicating that they originate from optical phonons.~\cite{Sichelschmidt2008}

%%%%%%%%%%%%%%%%  FIG.2  %%%%%%%%%%%%%%
\begin{figure}[t]
\begin{center}
\includegraphics[width=0.32\textwidth]{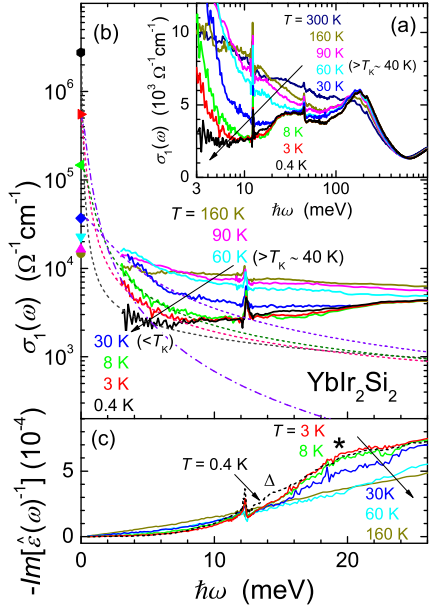}
\end{center}
\caption{
(Color online) Temperature dependence of the real part of optical conductivity $\sigma_1(\omega)$ spectrum (solid lines) in the logarithmic scale (a) and in the linear scale (b).
In (b), corresponding direct current conductivities ($\sigma_{DC}$, symbols) are also plotted.
Dashed lines: {\it Extended} Drude model (Eq.~\ref{tau2}) in which the scattering rate has the frequency dependencies shown in Fig.~3(b).
Dot-dashed line: {\it Normal} Drude model (Eq.~\ref{NormalDrude}) fitted to the $\sigma_1(\omega)$ spectrum at 3~K.
(c) Temperature dependence of the loss function $-{\rm Im}[\hat{\varepsilon}(\omega)^{-1}$] spectra.
Note the presence of the hump structure at 19~meV marked by an asterisk below $T_{\rm K}$ and the additional one at 14~meV marked by an open triangle at 0.4~K.
}
\label{OC}
\end{figure}
%%%%%%%%%%%%%%%%%
The $\sigma_1(\omega)$ and $-{\rm Im}[\hat{\varepsilon}(\omega)^{-1}$] spectra that were derived from the $R(\omega)$ spectra via KKA are displayed in Fig.~\ref{OC}.
Figure~\ref{OC}(a) shows $\sigma_1(\omega)$ over an enlarged energy range of 3--1000~meV for all temperatures used in this investigation.
As observed in the $R(\omega)$ spectrum, at 300~K and below 0.6~eV a Drude-type spectral shape dominates $\sigma_1(\omega)$ that monotonically increases with decreasing photon energy. 
This indicates that the $c$-$f$ hybridization intensity is relatively weak at 300~K, {\it i.e.}, the $4f$ electrons are localized.
With decreasing temperature, $\sigma_1(\omega)$ strongly decreases at the lowest energies, while at the same time, a shoulder structure appears at 25~meV below 30~K ($< T_{\rm K}$).
The peak at 200~meV and the shoulder at 25~meV can be roughly explained by the LDA band structure with the renormalization factor of 0.42 for the Yb~$4f$ state.~\cite{Kimura2009} 
However, a part of the shoulder at 25~meV may originate from the $c$-$f$ hybridization gap because, 
(i) below $T_{\rm K}$, the shoulder becomes clear below $T_{\rm K}$ and larger than the calculated intensity, and, 
(ii) as will be discussed further below,  the shoulder can be influenced by a magnetic field.
Moreover, decreasing the temperature also leads to a considerable shift of the Drude weight towards lower energies. 
For $T<T_{\rm K}$, the direct current conductivity ($\sigma_{DC}$, symbols in Fig.~\ref{OC}(b)) drastically increases while, at around 5~meV, $\sigma_1(\omega)$ is strongly decreasing. 
In contrast, for $T>T_{\rm K}$, the $\sigma_1(\omega,T)$ intensity at the lowest energies is similar to $\sigma_{DC}(T)$. 
This indicates that the Drude peak due to the heavy quasiparticles becomes sharp with decreasing temperature ($T<T_{\rm K}$).

Note that the $\sigma_1(\omega)$ spectra cannot be fitted by a normal Drude formula [$\sigma_1(\omega)=\sigma_{DC}/(1+\omega^2\tau^2)$, a dot-dashed line in Fig.~\ref{OC}(b)] as already reported for YbRh$_2$Si$_2$.~\cite{Kimura2006}
An {\it extended} Drude formula is plotted by dashed lines in Fig.~\ref{OC}(b) providing a reasonable description of the low-energy $\sigma_1(\omega)$ spectra. 
As will be discussed later, this extended Drude formula contains a scattering rate $1/\tau$ with a power-law dependence on $\hbar\omega$. 

The loss function spectra in Fig.~\ref{OC}(c) show characteristic hump structures marked by an asterisk and an open triangle. 
The one at 19~meV appears below $T_{\rm K}$ and therefore can be attributed to the plasmon peak of the heavy quasiparticles. 
At 0.4~K, the hump intensity at 19~meV slightly decreases and an additional hump appears at 14~meV (open triangle). 
Because also the electronic specific heat coefficient $\gamma$ is constant at $T\le$~0.4~K, we assign the 14~meV peak in the loss function to the plasmon peak of the quasiparticles in the Fermi liquid state.

%%%%%%%%%%%%%%%  FIG.3  %%%%%%%%%%%%%%%
\begin{figure}[t]
\begin{center}
\includegraphics[width=0.35\textwidth]{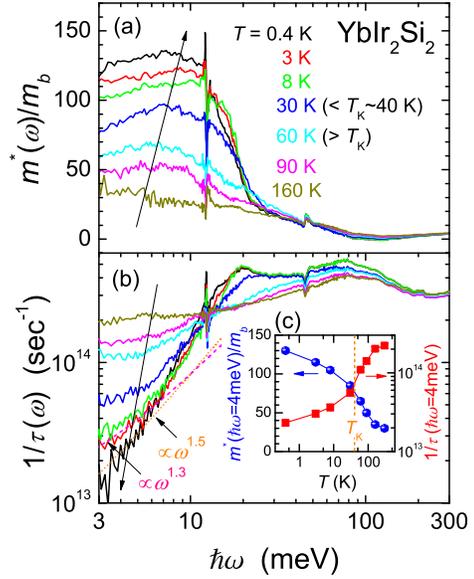}
\end{center}
\caption{
(Color online) Temperature dependence of (a) the effective mass relative to the band mass, $m^{*}(\omega)/m_b$, and (b) the scattering rate $1/\tau(\omega)$ as a function of photon energy $\hbar\omega$.
Dotted and dashed lines emphasize $1/\tau\propto\omega^{1.5}$ and $1/\tau\propto\omega^{1.3}$ behaviors for $T$~=~0.4~K and 3~K, respectively.
(c) Temperature dependence of $m^{*}(\omega)/m_b$ and $1/\tau(\omega)$ at $\hbar\omega$~=~4~meV.
The Kondo temperature $T_{\rm K}$ is also shown.
}
\label{ExtendedDrude}
\end{figure}
%%%%%%%%%%%%%%%%%%%%%%%%%%%%%%
In order to clarify the photon energy and temperature dependences of the quasiparticle scattering rate, an {\it extended} Drude analysis in terms of an energy dependence of the effective mass [$m^*(\omega)$] and scattering rate [$1/\tau(\omega)$] was performed.
As shown in Fig.~\ref{ExtendedDrude}, $m^*(\omega)/m_b$ gradually increases below 20~meV, and $1/\tau(\omega)$ gradually decreases below 10~meV.
The temperature dependence of this behavior is illustrated in Fig.~\ref{ExtendedDrude}(c) for a representative energy $\hbar\omega$~=~4~meV. 
Both $m^*/m_b(T)$ and $1/\tau(T)$ continuously change with temperature. 
However, at $T_{\rm K}\sim$40~K, the change occurs most rapidly that shows an enhanced formation of heavy quasiparticles when cooling to below $T_{\rm K}$.
This is consistent with a conventional view of the evolution of heavy fermions, but it is an important observation in $\sigma_1(\omega)$ spectra.
Note that $m^*(\omega)/m_b$ below 10~meV is similar to that of a related material YbRh$_2$Si$_2$~\cite{Kimura2006} indicating similar mass enhancement of heavy fermion states in these materials.

The dotted and dashed lines in Fig.~\ref{ExtendedDrude}(b) indicate power-law dependences of ($1/\tau(\omega)$) in the low energy region and at  temperatures of 0.4~K and 3~K which are well below $T_{\rm K}$. 
We fitted
\begin{equation}
\frac{1}{\tau(\omega,T)} = \frac{N e^2 \rho(T)}{m_b}+b^* (\hbar\omega)^{n(T)}
\label{tau}
\end{equation}
to the data, where $\rho(T)$ is the zero energy DC electrical resistivity and $b^*$ is a constant related to $b$ in Eq.~\ref{tau1}.
The fit parameters for $T=0.4$ and 3~K, are $1/\tau(\omega)=2.5\cdot10^{12}+8.0\cdot10^{16}\cdot(\hbar\omega)^{1.5}$, and $1/\tau(\omega)=1.0\cdot10^{13}+2.9\cdot10^{16}\cdot(\hbar\omega)^{1.3}$, respectively. 
Equation~\ref{tau} describes the data well below 5.5~meV for 0.4~K and below 7~meV for 3~K. 
For the sake of clarity, we refrained from showing data fits for $T$~=~0.4 and 3~K in Fig.~\ref{ExtendedDrude}(b). 
These results indicate that the photon energy dependence of the scattering rate follows a similar power-law behavior as the temperature dependence of $\rho(T)$: $\rho(T)\propto T^{1.5}$ at 0.4~K and $\rho(T)\propto T^{1.3}$ at 3~K \cite{Hossain2005}. 
From this comparison, we infer that a NFL behavior is also indicated in the optical response of heavy quasiparticles for photon energies up to at least 5~meV.
Note that the $1/\tau(\omega)$ spectrum of YbRh$_2$Si$_2$ below 10~K is proportional to $\omega^1$ with a same power-law dependence on temperature in the photon energy region below 7~meV higher $\hbar\omega$ than that of YbIr$_2$Si$_2$.
This may indicate that the NFL energy region of YbRh$_2$Si$_2$ is wider than that of YbIr$_2$Si$_2$.
 
Using above results for the energy dependencies of $1/\tau(\omega)$, we also employed an extended Drude formula to describe the spectra in Fig.~\ref{OC}(b) with
\begin{equation}
\sigma_1(\omega) = \sigma_{DC}\frac{\tau(\omega)}{\tau(0)}\frac{1}{1+\omega^2\tau(\omega)^2}.
\label{tau2}
\end{equation}
As shown by the dashed lines in Fig.~\ref{OC}(b), Eq.~\ref{tau2} describes the $\sigma_1(\omega)$ data reasonably well in the same energy region as $1/\tau(\omega)$. 
This indicates that the Drude peak of heavy quasiparticles contains a scattering rate which shows the same power-law dependency on the photon energy as the temperature power-law dependency of the electrical resistivity. 
Moreover, this justifies the use of an extended Drude description of $\sigma_1(\omega)$. 
Note that this result is not inconsistent with the finding that the microwave conductivity of heavy quasiparticles obeys a normal Drude formula,~\cite{Scheffler2005} because $\sigma_{DC}$ is dominant in a sufficiently low energy region.

%%%%%%%%%%%%%%%  FIG.4  %%%%%%%%%%%%%%%
\begin{figure}[t]
\begin{center}
\includegraphics[width=0.32\textwidth]{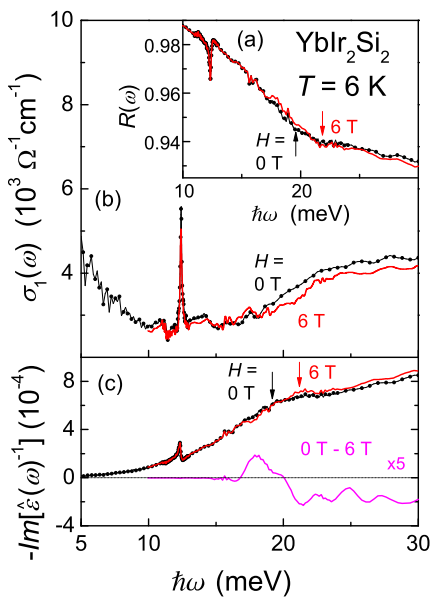}
\end{center}
\caption{
(Color online) Magnetic field ($H$) dependence of the reflectivity $R(\omega)$ (a), the optical conductivity $\sigma(\omega)$ (b), and the loss function $-Im[\hat{\varepsilon}(\omega)^{-1}]$ (c) spectra of YbIr$_2$Si$_2$ at $T$~=~6~K.
Arrows indicate the plasma edges of heavy quasiparticles observed at $H$~=~0 and 6~T.
The subtraction spectrum of $-Im[\hat{\varepsilon}(\omega)^{-1}]$ at 6~T from that at 0~T is also plotted.
}
\label{Hdep}
\end{figure}
%%%%%%%%%%%%%%%%%%%%%%%%%%%%%%

Finally, we investigated the behavior of the hump structure in the energy loss spectra (see asterisk in Fig.~\ref{OC}(c)) in the presence of a magnetic field. 
If this hump is related to a heavy quasiparticle plasmon a magnetic field should shift it towards higher energies concomitantly with the reduction of the mass enhancement.~\cite{Dordevic2006}
Figure~\ref{Hdep} shows the magnetic-field-dependent $R(\omega)$ (a), $\sigma_1(\omega)$ (b), and $-{\rm Im}[\hat{\varepsilon}(\omega)^{-1}]$ (c) spectra at 6~K. 
Indeed, by applying a magnetic field of 6~T, characteristic changes are observable as indicated by the arrows in Figs.~\ref{Hdep}(a) and (c). 
These changes are fairly small but they are outside the experimental error (less than 0.2~\% accuracy in $R(\omega)$) and can be seen most clearly in the difference energy loss spectra in Fig.~\ref{OC}(c). 
The plasma edge ($\hbar\omega_p$) of the heavy quasiparticles in the $R(\omega)$ spectrum as well as the hump in the $-{\rm Im}[\hat{\varepsilon}(\omega)^{-1}]$ spectrum slightly shift to the higher energy side from $19.3\pm0.3$~meV to $20.6\pm0.3$~meV. 

The relation between the magnetic-field-dependent effective mass $m^*(H)$ and $\hbar\omega_p(H)$ is $m^*(H)/m^*(H=0)=[\hbar\omega_p(H=0)/\hbar\omega_p(H)]^2$ and, hence, the effective mass at 6~T is suppressed by $12\pm5\%$ from the zero-field value.
Within a mean field theory for the hybridization, the field dependence of the quasiparticle effective mass proportional to $1-3/2(H/H_0)^2$ with a representative magnetic field $H_0$.~\cite{Beach2005}
With our results for the suppression of the effective mass, we obtain $H_0=21\pm9$~T. 
The $H_0$ value is similar to the Kondo energy of YbIr$_{2}$Si$_{2}$ in magnetic units is $H_{\rm K}=18$~T ($g_{\perp}=3.35$~\cite{Gruner2010} and $T_{\rm K}=40$~K).
This result is not consistent with that of a heavy fermion material CeRu$_4$Sb$_{12}$~\cite{Dordevic2006}, {\it i.e.}, $H_0$ is close to the coherence temperature $T^*\sim$50~K lower than $T_{\rm K}\sim$100~K.
In our result, the similar values of $H_{0}$ and $H_{\rm K}$ imply that the effective mass decreases with applied magnetic field similar to what is known for $T_{\rm K}$.~\cite{Beach2005}
This signature of the Kondo effect leads to a $c$-$f$ hybridization gap near 20~meV in YbIr$_{2}$Si$_{2}$.

%------------------------------------------------------------------------------
%
%\section{Conclusion}
%
In conclusion, the temperature and magnetic field dependencies of reflectivity as well as optical conductivity spectra of YbIr$_2$Si$_2$ were measured and the dynamical properties of heavy quasiparticles were discussed.
At $T=0.4$~K below photon energies of 15~meV, the effective mass is enhanced by a factor of 130 compared to the unrenormalized band mass.
Below the Kondo temperature, the scattering rate as a function of photon energy obeys a similar power-law behavior as the temperature dependence of the electrical resistivity. 
The effective mass is suppressed by $12\pm5\%$ from the zero-field value with increasing magnetic field to 6~T.
The magnetic field dependence of the effective mass is inversely proportional to the Kondo temperature.

%%%%%%%%%%%%%%%%%%%%%%%%%%%%%%
%\section*{Acknowledgments}
Part of this work was performed by the Use-of-UVSOR Facility Program (BL7B, 2008) of the Institute for Molecular Science.
The work performed in Okazaki was partly supported by a Grant-in-Aid for Scientific Research (B) from MEXT of Japan (Grant No.~22340107), and part of the work done in Dresden was supported by the DFG Research Unit 960 ``Quantum Phase Transitions''.
%
%%%%%%%%%%%%%%%%%%%%%%%%%%%%%%

\end{document}